# On the power counting of loop diagrams in general relativity

John F. Donoghue* and Tibor Torma†

*University of Massachusetts, Department of Physics, Amherst MA 01003*

August 13, 2013

**Abstract**

A class of loop diagrams in general relativity appears to have a behavior which would upset the utility of the energy expansion for quantum effects. We show through the study of specific diagrams that cancellations occur which restore the expected behaviour of the energy expansion. By considering the power counting in a physical gauge we show that the apparent bad behavior is a gauge artifact, and that the quantum loops enter with a well behaved energy expansion.

*e–mail: donoghue@phast.umass.edu
†e–mail: kakukk@phast.umass.edu

# I Introduction

Loop calculations in general relativity are readily interpreted using the techniques of effective field theory [1, 2]. As in all effective field theories, the utility of such calculations is tied to an expansion in powers of the energy or inverse distance. In chiral theories, Weinberg [3] has provided an important theorem which states that diagrams with increasing numbers of loops contribute to an amplitude with increasing powers of the energy, with each extra loop adding an extra factor of $E^2$. For example, if one is working to order $E^4$ accuracy one needs to include only one loop diagrams. While pure gravity behaves exactly in the same way, if we try a simple extension of this same argument to gravity interacting with matter, we will see in Sec. II that the desired behavior is not obtained. There is a class of diagrams which appears to have $Gm^2$ as the expansion parameter. This would upset the utility of the energy expansion. The purpose of this paper is to explore this problem and to see if it obstructs the energy expansion.

The desired expansion parameters for quantum corrections in an effective theory of gravity is $Gq^2 \sim \frac{G}{r^2}$, such that at low energies/long distances the higher order loop effects are suppressed with respect to tree diagrams and low order loops. Thus we can obtain predictions to a given order with a finite amount of calculation. General relativity also contains the classical expansion parameter $Gmq \sim \frac{Gm}{r}$ which represents the nonlinearities of the classical theory. This can be found in the loop expansion from the nonanalytic terms of the form $Gq^2 \sqrt{\frac{m^2}{-q^2}}$. However, $Gm^2$ as an expansion parameter is a major problem. In the first place, the mass can be extremely large in units of the Planck mass (e. g. $m = M_{\text{Sun}}$) so that $Gm^2$ can be a number very much larger than unity. In addition if we restore factors of $\hbar$, this dimensionless combination goes like $\frac{Gm^2}{\hbar}$. The classical limit $\hbar \to 0$ would be upset by corrections of this form.

We will see that the apparent difficulty with the loop expansion appears to be a gauge artifact. When calculating in harmonic gauge, where the power counting is first discussed in Sec. II, there occur cancellations between individual diagrams, cancelling the bad behavior. We detail the calculation for the box and crossed box diagrams. Part of the problem is due to the occurrence of both classical and quantum effects in the same Feynman diagram, when treated in covariant gauges. This suggests that separating the classical physics from the physical quantum (transverse and traceless) degrees of freedom will improve the power counting. For the interaction of two nearly static masses, we show that this is in fact the case.

The organization of this paper is as follows. In Sec. II we make a naïve generalization of the Weinberg power counting theorem and isolate those diagrams which appear to give a problem. In order to explore this without all the tensor indices of gravitons, we introduce a scalar toy model with the same behavior in Sect. III in order to see how cancellations occur. Sec. IV applies the lessons so learned to the gravitational interaction. Sec. V was devoted to development of the power counting scheme in a physical gauge, and to the interpretation of the apparent problem as a gauge artifact. We end with some concluding comments.

## II  Power counting in covariant gauges

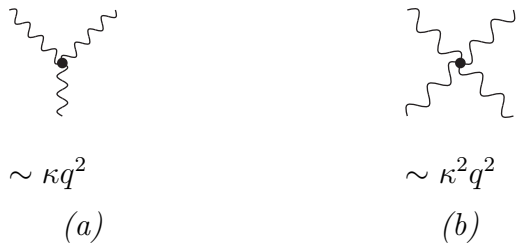

Figure 1: Three and four graviton couplings.

We are interested in treating powers of energies and masses in vertices and propagators in order to determine the overall energy dependence of a given multiloop diagram. The mass of the matter field is not a small parameter, but we can treat the external three-momenta as small if we are working at low enough energies. Let us review the Feynman rules, and extract the essential dependence of the vertices. Starting from the action

$$S = \int d^4x \sqrt{g} \cdot \frac{2}{\kappa^2} \cdot R \tag{II.1}$$

with $\kappa^2 = 32\pi G$, we expand this metric

$$g_{\mu\nu} = \eta_{\mu\nu} + \kappa h_{\mu\nu} \tag{II.2}$$

where $h_{\mu\nu}$ is the fluctuating field. Expanding $\frac{2}{\kappa^2}R$ in powers of $h_{\mu\nu}$ we see that a term which involves $n$ graviton fields, i. e. $h^n$, carries a coupling constant $\kappa^{n-2}$. Since the curvature is second order in derivatives, all terms emerging from the Einstein action will be of order $q^2$. Thus the triple graviton coupling of Fig. 1 is of order $\kappa q^2$, while the four graviton vertex is of order $\kappa^2 q^2$ etc. The matter fields couple to gravitons through $T_{\mu\nu}$, which for a scalar field has matrix elements

$$< p' \mid T_{\mu\nu} \mid p > = p_\mu p'_\nu + p_\nu p'_\mu - \frac{1}{2} g_{\mu\nu} \left( p \cdot p' - m^2 \right) \tag{II.3}$$

with $p_\mu = \left( \sqrt{m^2 + \vec{p}^2} \mid \vec{p} \right)_\mu$. Treating the mass as a large parameter leads to a one graviton vertex (see Fig. 2) which behaves as $\kappa m^2$ while the two graviton diagram is of order $\kappa^2 m^2$ etc.

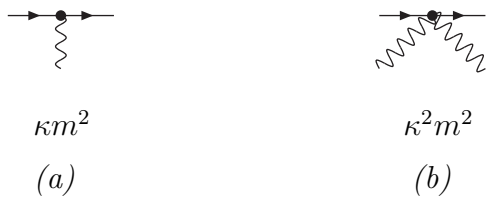

$\kappa m^2$ $\quad\quad\quad\quad\quad\quad\quad\quad$ $\kappa^2 m^2$

(a) $\quad\quad\quad\quad\quad\quad\quad\quad\quad$ (b)

Figure 2: Matter-graviton couplings from $T_{\mu\nu}$.

The graviton propagator, like all massless boson propagators, scales as $\frac{1}{q^2}$. The matter field propagator requires a bit more explanation. Because we are dealing with an effective theory at low energies, we need not consider loops of heavy matter fields. These loops have already been integrated out in order to define the low energy effective theory. However, we do need to consider matter fields which appear as external states and which propagate through a given diagram interacting with each other and with gravitons. The explicit form of the propagator is

$$D(p+q) = \frac{i}{(p+q)^2 - m^2} = \frac{i}{2p \cdot q + q^2 + (p^2 - m^2)} \tag{II.4}$$

where $p$ is the momentum that the matter field has as an external particle, and $q$ is the momentum which has been added to it through interactions with gravitons (internal or external). The external momentum is on shell ($p^2 - m^2 = 0$) so that the matter propagator is counted as a factor of $\frac{1}{mq}$. Note that if we had chosen a different normalization for our matter fields (e. g. a nonrelativistic normalization such that $T_{00} \sim m$ and $D(q) \sim \frac{1}{q}$) both the vertices and propagators change in a way that compensates each other, leading to the same counting rules as in our normalization.

Before giving the general power counting theorem, let us illustrate the idea with two specific examples, one of which illustrates the "good" behavior and one which shows the problem. First consider graviton-graviton scattering, whose overall matrix element is dimensionless. At lowest order we have a $\frac{1}{\kappa^2}$ factor from the coupling in the Einstein action, one of $\kappa^4$ from the four graviton fields and one $q^2$ because the Einstein action involves two derivatives. This leads to an overall matrix element

$$\mathcal{M}_{tree} \sim \kappa^2 q^2. \tag{II.5}$$

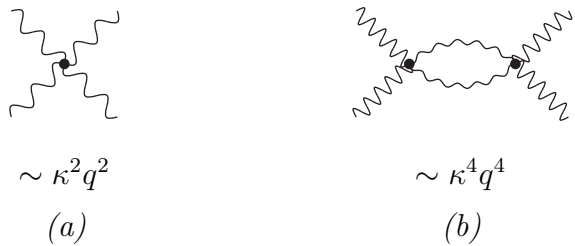

$\sim \kappa^2 q^2$ $\qquad\qquad\qquad \sim \kappa^4 q^4$

(a) $\qquad\qquad\qquad\qquad$ (b)

Figure 3: Sample four-graviton interaction diagrams to one loop, illustrating the expected behavior in the energy expansion.

If we try to iterate this vertex to produce the one loop diagram of Fig. 3b we obtain schematically

$$\mathcal{M}_{loop} \sim \kappa^4 \int \frac{d^4 l}{(2\pi)^4} \frac{(l-p_1)^2 (l-p_2)^2}{l^2 (l-q)^2} \tag{II.6}$$

where $p_1, p_2, q$ are various combination of external momenta. If this loop integral is regularized dimensionally, which does not introduce powers of any new scale, the integral will be represented in terms of the exchanged momentum to the appropriate power. Thus we have

$$\mathcal{M}_{loop} \sim \kappa^4 q^4 \tag{II.7}$$

where again $q$ represents some combination of external momenta. [There may also be logarithms of $\frac{q^2}{\mu^2}$ where $\mu$ is the usual scale introduced in dimensional regularization.] In this case adding a loop has generated an effect which is higher order in the energy expansion. The expansion is in terms of powers of $\kappa^2 q^2$.

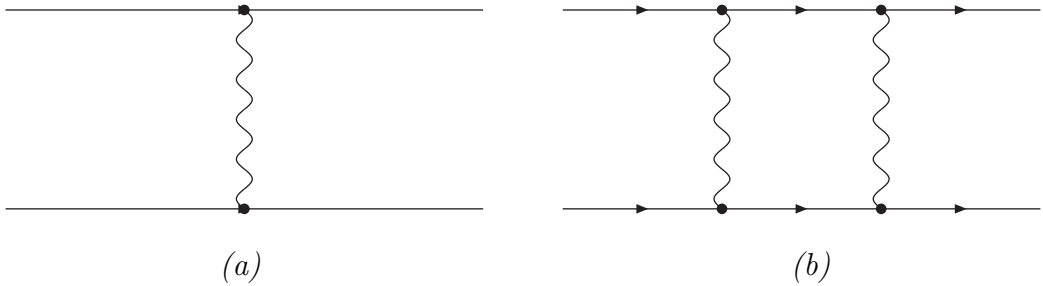

(a)     (b)

Figure 4: Sample interactions of two massive particles.

A different behavior is shown by the interactions of two massive particles, such as in Fig. 4a,b. The tree level result in our normalization is

$$\mathcal{M}_{tree} = \kappa^2 \cdot \frac{m_1^2 m_2^2}{q^2} \tag{II.8}$$

which again is dimensionless. Iterating this to form a loop gives us

$$\mathcal{M}_{loop} \sim \kappa^4 m_1^4 m_2^4 \cdot \int d^4 l \cdot \frac{1}{m_1(l+p)} \cdot \frac{1}{m_2(l+p')} \cdot \frac{1}{(l+q')^2} \cdot \frac{1}{(l+q)^2} \tag{II.9}$$

which by the same reasoning is

$$\mathcal{M}_{loop} \sim \kappa^4 \cdot \frac{m_1^3 m_2^3}{q^2} \sim \kappa^2 \cdot \frac{m_1^2 m_2^2}{q^2} \cdot \kappa^2 m_1 m_2. \tag{II.10}$$

Here the expansion parameter appears as $\kappa^2 m^2$. An explicit calculation of this diagram later in this paper confirms that this is the correct result for the diagram by itself. This expansion parameter $\kappa^2 m^2$ would cause the problem described in the introduction.

Now let us turn to the general power counting result. Our goal is to obtain the power of $q$ (with $q$ being a typical external momentum) that a general diagram would yield. This will tell us what order in the energy expansion that diagram will contribute to. The problematic class will be manifest by having diagrams with increasing number of loops which yield the same power of $q$, so that to calculate to this order in the energy expansion one would apparently needs to sum all the diagrams in this class. For a general result we need to allow for vertices not just from the lowest order gravitational action, but also from ones which contain more derivatives. Let us write this schematically as

$$S_g = \int d^4x \sqrt{g} \cdot \frac{2}{\kappa^2} \cdot \left[ R + \kappa_0^2 R^2 + \kappa_0^4 R^3 + \ldots \right] \quad \text{(II.11)}$$

such that the coefficients of a gravitational Lagrangian with $n$ derivatives will be $\frac{\kappa_0^{n-2}}{\kappa^2}$. Note that $\kappa_0 \sim \frac{1}{\text{energy}}$. In a pure gravitational theory one would expect $\kappa_0 \sim \kappa$, but there is no need to impose such a restriction here. Likewise the matter Lagrangian can involve extra derivatives on the light fields. We let the coefficients of the higher derivative terms involve a scale $\overline{\kappa_0}$, i. e.

$$\begin{aligned} S_m &= \int d^4x \sqrt{g} \cdot \\ &\quad \cdot \left[ \frac{1}{2} \left( \partial_\mu \Phi \partial^\mu \Phi - m^2 \Phi^2 \right) + \overline{\kappa_0}^2 R \partial_\mu \Phi \partial^\mu \Phi + \kappa_0^4 R^2 \partial_\mu \Phi \partial^\mu \Phi + \ldots \right] \end{aligned} \quad \text{(II.12)}$$

so that the coefficient of a Lagrangian with $l$ derivatives on the gravitational field is $m^2 \overline{\kappa_0}^l$. [Again, $\overline{\kappa_0} \sim \frac{1}{\text{energy}}$ and $\overline{\kappa_0}$ can be kept distinct from $\kappa_0$ and $\kappa$ if desired.]

Our procedure is to count powers of $\kappa, \kappa_0, \overline{\kappa_0}$ and $m^2$ in a general diagram. The remaining energy factor of the diagram, needed to give the proper overall dimension, will come from factors of the external momenta. Consider a diagram with $N_E^m$ external matter legs and $N_E^g$ external graviton legs, with a series of interactions between these particles. Correspondingly let $N_I^m$ and $N_I^g$ be the number of internal matter and graviton propagators respectively. There are $N_V^g$ vertices involving only gravitons, and $N_V^m$ vertices which involve matter fields plus any number of gravitons, and a total of $N_L$ loops. However, these vertices need to be categorized by the number of derivatives that are involved. For example, let $N_V^g[n]$ be the number of graviton vertices which come from a Lagrangian with $n$ derivatives. Clearly, $N_V^g = \sum_n N_V^g[n]$. Likewise the number of matter vertices with $l$ derivatives on the light fields will be called $N_V^m[l]$ with $N_V^m = \sum_l N_V^m[l]$. We illustrate this with a sample

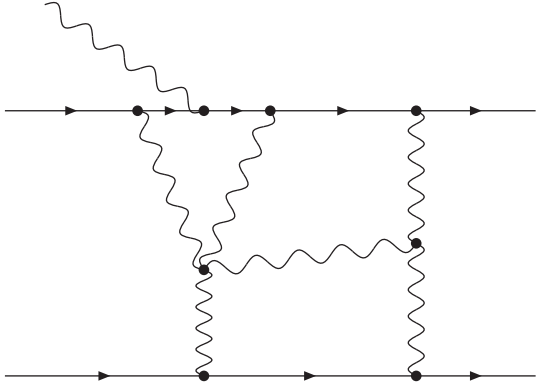

Figure 5: Sample diagram with $N_E^g = 1$, $N_E^m = 4$, $N_I^g = 6$, $N_I^m = 4$, $N_V^g = 2$, $N_V^m = 6$, $N_L = 3$.

diagram in Fig. 5. All matter lines propagate all the way through a diagram without terminating.

With these definitions the coupling constants contribute the dimensionful factors

$$\left(\kappa^2\right)^{-N_V^g} \cdot \left(\kappa_0\right)^{\sum_n (n-2) N_V^g[n]} \cdot \left(m\right)^{2 N_V^m} \cdot \left(\overline{\kappa_0}\right)^{\sum_l l \cdot N_V^m[l]}. \tag{II.13}$$

In addition, because each internal graviton line is formed using two vertices, the graviton fields will contribute a power of

$$\left(\kappa\right)^{2 N_I^g + N_E^g} \tag{II.14}$$

from the normalization of the metric in (II.2). On a matter line, there will be $(v-1)$ propagators if there are $v$ vertices. Thus the number of matter propagators $N_I^m$ satisfies

$$N_I^m = N_V^m - \frac{1}{2} N_E^m. \tag{II.15}$$

Since each propagator counts as a power of $\frac{1}{m}$, this contributes mass factors

$$\left(\frac{1}{m}\right)^{N_V^m - \frac{1}{2} N_E^m}. \tag{II.16}$$

These constitute all of the general dimensionful parameters except the external momenta and the loop momenta. When the loop integrals are regularized dimensionally there will not be any powers of a regulator mass and the remaining dimensions after integration will be carried by the external momenta. Let us generically call these momenta $q$, and describe the power of the momenta by a factor $q^D$. It is the dimension $D$ which we are seeking in this exercise.

Overall, this matrix element carries a dimension

$$\mathcal{A} \sim \left(\text{Energy}\right)^{4-N_E^m - N_E^g}. \tag{II.17}$$

From our identification above, this is decomposed as

$$\begin{aligned}\mathcal{A} &\sim \left(\text{Energy}\right)^{4-N_E^m - N_E^g} \\ &= \left(\kappa_0\right)^{\sum_n (n-2)\cdot N_V^g[n]} \cdot \left(m\right)^{2N_V^m} \cdot \left(\overline{\kappa_0}\right)^{\sum_l l \cdot N_V^m[l]} \cdot \left(\kappa\right)^{2N_I^g + N_E^g - 2N_V^g} \cdot \\ &\quad \cdot \left(\frac{1}{m}\right)^{N_V^m - \frac{1}{2}N_E^m} \cdot q^D.\end{aligned} \tag{II.18}$$

There are however some relations among all the variables. For example, the total number of internal lines can be expressed in terms of the total number of vertices and the number of loops. The relation is

$$\begin{aligned}N_I^m + N_I^g &= N_L + (N_V^m + N_V^g) - 1 \\ &= N_L + \sum_l N_V^m[l] + \sum_n N_V^g[n] - 1.\end{aligned} \tag{II.19}$$

We can use this to eliminate $N_I^g$, using also $N_I^m = N_V^m - \frac{1}{2}N_E^m$, to find

$$\begin{aligned}N_I^g &= (N_I^g + N_I^m) - \left(N_V^m - \frac{1}{2}N_E^m\right) \\ &= N_L + \frac{1}{2}N_E^m + \sum_n N_V^g[n] - 1.\end{aligned} \tag{II.20}$$

Plugging this into the general formula Eq. (II.18), using $\sum_n N_V^g[n] = N_V^g$ and recalling that $\kappa, \kappa_0, \overline{\kappa_0}$ all go as $\frac{1}{\text{Energy}}$ allow us to solve for the parameter $D$, resulting in

$$D = 2 - \frac{N_E^m}{2} + 2N_L - N_V^m + \sum_n (n-2) N_V^g[n] + \sum_l l \cdot N_V^m[l]. \tag{II.21}$$

This is our general power counting result for the momentum dependence of a general diagram. If we disregard the matter vertices, $N_E^m = N_V^m[l] = N_V^m = 0$, it is identical to Weinberg's theorem for chiral theories. The momentum dimension of a diagram is higher if we increase the number of loops or if we use a gravitational Lagrangian with more than two derivatives. This shows that the power counting of loop diagrams in pure gravity involves the parameter $\kappa^2 q^2$ (or $\kappa_0^2 q^2$ if $\kappa_0 \neq \kappa$.)

In the presence of matter, the last term also behaves as expected: using the Lagrangian with extra derivatives on light fields ($l > 0$) only increases the power of the momentum. However, the problem arises because of the minus sign in front of $N_N^m$. Increasing the number of matter vertices in diagrams does not increase the order in the energy expansion. This cannot actually make the momentum power $D$ decrease by increasing the number of matter vertices, since to add matter vertices to a given process we also have to change the number of loops. However, there are diagrams where one can increase the number of loops by one while increasing the number of matter vertices by two. Fig. 4 is one such example. This leaves $D$ unchanged. Thus higher loop processes contribute at the same level to the energy expansion as tree processes. This gives a loop expansion of $\kappa^2 m^2$ instead of $\kappa^2 q^2$.

In summary, we have computed the momentum power or a given process (II.13), and found a class of dangerous diagrams where the addition of two matter vertices adds only one loop to the process, leading to no net increase in the momentum power $D$.

## III  Analysis of a related model

Because the couplings of general relativity are complicated by the tensor indices, it is easier to analyse a simpler model first. Although it is clear that the model is not identical with general relativity, it will nevertheless exhibit several interesting features which we will be able to generalize to the case of gravity in the next section.

Consider a massless scalar field $h$ coupled to one or more massive scalars $\Phi$ with a trilinear coupling which carries the same strength as our counting rule given in (II.13). That is, we identify the Lagrangian

$$\mathcal{L} = \frac{1}{2}\partial_\mu \Phi \partial^\mu \Phi + m^2 \Phi^2 (1 + \kappa h) + \frac{1}{2}\partial_\mu h \partial^\mu h. \tag{III.1}$$

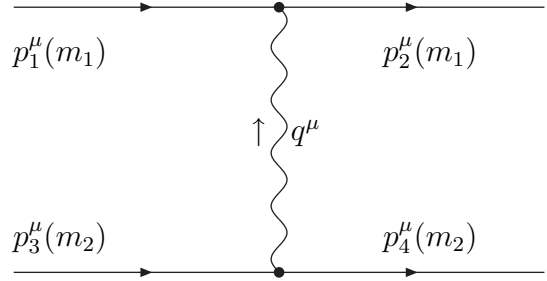

Figure 6: Tree level graph for heavy scalar scattering.

The coupling $\kappa m^2 \Phi^2 h$ enters into the power counting derivation in the same way as the lowest order graviton coupling and in this theory we obtain the same momentum power as in (II.13) with $N_V^g[n] = 0$ for $n > 2$ and $N_V^m[l] = 0$ for $l > 0$. The dangerous class of diagrams identified in the previous section also are equally problematic for this model.

Let us verify the result of the counting theorem by considerations of the gravitational interaction of two heavy masses, as in Fig. 6, 7a,b. The single "graviton" exchange vertex, Fig. 6, has magnitude

$$\mathcal{M} = \kappa^2 \frac{m_1^2 m_2^2}{q^2} \tag{III.2}$$

as expected. Note that in forming a nonrelativistic static potential one divides by $2m_1 \cdot 2m_2$ to account for our normalization of the states, obtaining

$$\begin{aligned} V(r) &= \int \frac{d^3 q}{(2\pi)^3} \cdot \frac{1}{2m_1 \cdot 2m_2} \cdot \kappa^2 \frac{m_1^2 m_2^2}{q^2} \cdot e^{i\vec{q}\cdot\vec{r}} \\ &= -\kappa^2 \cdot \frac{m_1 m_2}{32\pi \cdot r}. \end{aligned} \tag{III.3}$$

Now consider the box diagram. Fig. 7a:

$$\mathcal{M}_{box} = i \cdot (\kappa m_1^2)^2 \cdot (\kappa m_2^2)^2 \cdot \int \frac{d^4 k}{(2\pi)^4} \frac{1}{[k^2 + i\epsilon][(q-k)^2 + i\epsilon]} \cdot \\ \cdot \frac{1}{(p_1 + k)^2 - m_1^2} \cdot \frac{1}{(p_2 - k)^2 - m_2^2}. \tag{III.4}$$

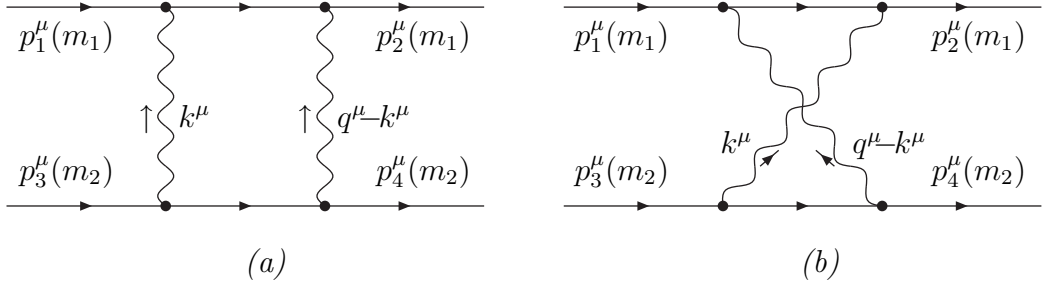

Figure 7: The (a) box and (b) crossed box graphs which have the wrong naïve power counting behavior.

We combine the graviton propagators using the usual identity

$$\frac{1}{k^2}\frac{1}{(k+q)^2} = \int_0^1 \frac{dx}{[k^2 - 2xk\cdot q + x^2 q^2]^2}$$
$$= \int_0^1 \frac{dx}{\left[(k-xq)^2 - \overline{Q}^2\right]} \quad \text{(III.5)}$$

with $\overline{Q}^2 = -x(1-x)q^2$, and do likewise for the matter propagator. The diagram is ultraviolet finite but has an infrared divergence which we regulate dimensionally. Integrating in $d$ dimensions we obtain

$$\mathcal{M}_{box} = -(\kappa m_1^2)^2 \cdot (\kappa m_2^2)^2 \cdot \mu^{4-d} \cdot \frac{\Gamma\left(4-\frac{d}{2}\right)}{(4\pi)^{\frac{d}{2}}} \cdot \int_0^1 dx \int_0^1 dy \int_0^1 dz \cdot \frac{y(1-y)}{D^{4-\frac{d}{2}}} \quad \text{(III.6)}$$

with

$$D = M_z^2 y^2 \overline{Q}^2 (1-y)^2 \quad \text{(III.7)}$$
$$M_z^2 = z^2 m_1^2 + (1-z)^2 m_2^2 - 2z(1-z) p_1 \cdot p_2.$$

For small $\frac{\overline{Q}}{m}$, the $y$ integration can be done near $d = 4$ dimensions,

$$\int_0^1 \cdot \frac{y(1-y)}{D^{4-\frac{d}{2}}} = \frac{1}{2M^2} \cdot \frac{1}{\overline{Q}^{6-d}} \cdot \left[1 + (d-4)\left(\frac{1}{2} - \frac{\overline{Q}}{m}\right)\right]. \quad \text{(III.8)}$$

The infrared divergence as $d \to 4$ is in the $x$ integration. In the $z$ integration there is a pole when the intermediate matter state is on shell and this is made well defined by the usual $i\epsilon$ factors. However, both of these integrals can be done straightforwardly. If we define $p_1 \cdot p_2 = m_1 m_2 + w$, we end up with

$$\mathcal{M}_{box} = \frac{\kappa^2 m_1^2 m_2^2}{q^2} \cdot \frac{\kappa^2 m_1 m_2}{16\pi^2} \cdot \left[ -1 + \frac{w}{3m_1 m_2} + i\pi \frac{m_1 m_2}{p(m_1 + m_2)} \right] \quad \text{(III.9)}$$
$$\cdot \left\{ \frac{2}{4-d} - \ln\left(-\frac{q^2}{m^2}\right) + \text{constant} + \mathcal{O}\left(\frac{q}{M}\right) \right\}$$

with $p = |\vec{p}|$ in the center of mass. If we defer comment on the imaginary part of this amplitude to below, we see that this does obey the expectation of the power counting theorem, with a correction of order $\kappa^2 m_1 m_2$ compared to the tree level amplitude.

However, we must also consider the crossed box diagram in Fig. 7b. This can be handled in a similar fashion, and is slightly easier because it does not have an imaginary part. Defining $p_1 \cdot p_4 = m_1 m_2 + w'$, we obtain

$$\mathcal{M}_{crossed} = \frac{\kappa^2 m_1^2 m_2^2}{q^2} \cdot \frac{\kappa^2 m_1 m_2}{16\pi^2} \cdot \left[ +1 - \frac{w'}{3m_1 m_2} \right] \cdot \quad \text{(III.10)}$$
$$\cdot \left\{ \frac{2}{4-d} - \ln\left(-\frac{q^2}{m^2}\right) + \text{constant} + \mathcal{O}\left(\frac{q}{M}\right) \right\}$$

with the *same* constant as in (III.9). We see that the most dangerous terms cancel between the diagrams. Using $w - w' = -\frac{q^2}{2}$ we get the final result

$$\mathcal{M}_{total} = \mathcal{M}_{tree} + \mathcal{M}_{box} + \mathcal{M}_{crossed} \quad \text{(III.11)}$$
$$= \frac{\kappa^2 m_1^2 m_2^2}{q^2} \cdot \left\{ 1 + \frac{1}{16\pi} \left[ -\frac{1}{6}\kappa^2 q^2 + i\pi \frac{\kappa^2 m_1 m_2}{p(m_1 + m_2)} \right] \cdot \right.$$
$$\left. \cdot \left[ \frac{2}{4-d} - \ln\left(-\frac{q^2}{m^2}\right) + \text{constant} \right] \right\}.$$

In the real part of the amplitude, the expansion parameter has become $\kappa^2 q^2$, which is well behaved. The imaginary part of the amplitude is just a phase and does not contribute to observables at this order when the matrix element is squared. It is simply the analogue of the well known "Coulomb phase" and is generated from the rescattering of the on shell intermediate state of matter particles. Like the Coulomb phase, it has been shown by Weinberg [4] to exponentiate to all orders in general relativity (the proof extends to this simpler theory as well), so that this term does not cause any trouble. The same paper by Weinberg also proves that infrared divergencies cancel in general relativity and by extension in this theory, by the considerations of virtual corrections of these diagrams plus the bremsstrahlung radiation of real particles. These can be regulated dimensionally also [5], and will yield as finite effects residual corrections of order $\kappa^2 q^2$ and $\kappa^2 q^2 \ln q^2$. We are not here directly interested in the exact answer; for us the important result was the cancellation of $\kappa^2 m^2$ effects in (III.11).

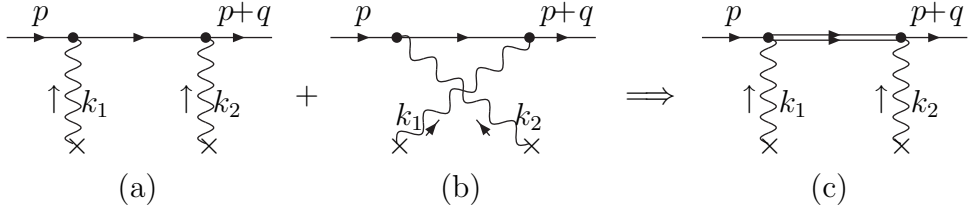

Figure 8: The definition of the 'Bose-symmetrized' propagators.

We can show that this cancellation is not peculiar to the box diagram, but rather is a general feature of this theory. This can be seen by considering the basic unit of a single line with two interactions with off-shell gravitons, as in Fig. 8. Any time a given ordering is possible, the crossed order is also possible, cf. Fig. 8a and 8b. If we add these two diagrams and allow the external legs to be on-shell, we find that the sum of propagators behaves as $\frac{1}{m^2}$ whereas each individual propagator was of order $\frac{1}{m}$:

$$\begin{aligned}
\mathcal{V} &= \kappa m^2 \cdot \left[ \frac{1}{(p+k)^2 - m^2} + \frac{1}{(p'-k)^2 - m^2} \right] \cdot \kappa m^2 \\
&= (\kappa m^2)^2 \cdot \left[ \frac{1}{2p \cdot k + k^2} + \frac{1}{-2p' \cdot k + k^2} \right] \cdot \quad \text{(III.12)} \\
&= (\kappa m^2)^2 \cdot \left[ \frac{2(p'-p) \cdot k - 2k^2}{[2p \cdot k + k^2] \cdot [2p' \cdot k - k^2]} \right].
\end{aligned}$$

However, since $p' - p = q$, there is no factor of the large mass $m$ in the numerator,

$$\mathcal{V} = (\kappa m^2)^2 \cdot \left[ \frac{q^2 - k^2 - (k-q)^2}{[2p \cdot k + k^2] \cdot [2p' \cdot k - k^2]} \right]. \quad \text{(III.13)}$$

Because there are two factors of $p \cdot k \sim m$ in the denominator, this double vertex counts as

$$\mathcal{V} \sim \frac{\kappa^2 m^4}{m^2} \sim \kappa^2 m^2 \tag{III.14}$$

rather than the $\mathcal{V} \sim \kappa^2 m^3$ that the naïve counting would imply. For Fig. 8a or b individually would give the extra two factors of $\frac{1}{m}$ which converts the undesirable expression to $\kappa^2 q^2$, thereby explaining the result found above.

The only exceptions to this power counting occurs for what can be termed "exceptional momenta." This refers to momenta where the propagator is not of order $\frac{1}{m}$, and can occur when the intermediate line goes on shell, $(p+k)^2 - m^2 = 0 = 2p \cdot k + k^2$ so that $p \cdot k \sim k^2$. In this case we do not gain a power the power of $\frac{1}{m}$ from the propagator. This is exactly what was found above in the explicit calculation of the box diagram. The on-shell intermediate states generate the imaginary part of the diagram which has a different dependence on the masses than does the remainder of the diagram. This leads to the Coulomb phase in the box diagram.

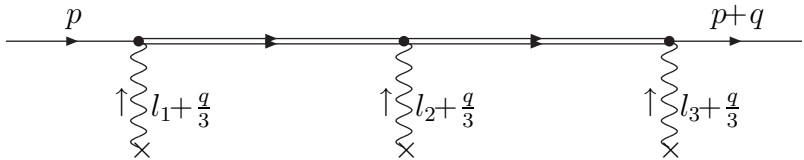

Figure 9: The two-loop 'Bose-symmetrized' propagator, defined for $\sum_1^3 l_i \equiv 0$.

We have been able to extend the demonstration of cancellations to three and four vertices (using computer algebra). The difficulty here appears because only the external lines are on shell. The desired cancellation does not occur for just the symmetrized sum of any two of the permutations, e. g. when we permute $l_1$ and $l_2$ in Fig. 9, but requires the sum of all six permutations.

We take the 'Bose-symmetrized propagator' denoted by a double line in Fig. 9:

$$D_m^{(3)}(p, l_1, l_2, l_3) = \frac{1}{3!} \cdot \sum_{\text{perm}} \frac{i}{(p + l_1 + \frac{q}{3})^2 - m^2} \cdot \frac{i}{(p + l_1 + l_2 + 2\frac{q}{3})^2 - m^2} \tag{III.15}$$

where the summation goes for all permutations of the $l_i$, subject to $\sum_i l_i \equiv 0$.

Naïve power counting would say $D^{(3)} \sim O\left(\frac{1}{m^2 q^2}\right)$ and for the total disappearance of the $\kappa m$ expansion parameter we would need $D^{(3)} \sim O\left(\frac{1}{m^4}\right)$ because the expected behavior of the amplitude is

in power counting:

$$\frac{(\kappa m_1^2)(\kappa m_2^2)}{q^2} \cdot \left\{1 + (\kappa m)^2_{\text{one-loop}} + (\kappa m)^2_{\text{two-loop}} + \ldots\right\} \tag{III.16}$$

and in the effective theory

$$\frac{(\kappa m_1^2)(\kappa m_2^2)}{q^2} \cdot \left\{1 + (\kappa q)^2_{\text{one-loop}} + (\kappa q)^2_{\text{two-loop}} + \ldots\right\}. \tag{III.17}$$

A tedious but straightforward algebra shows indeed that, disregarding special points in the integration region where some of the denominators are less than $O(m_i)$, cancellations occur with the result

$$D^{(3)}_m(p, l_1, l_2, l_3) = O\left(\frac{1}{m^4}\right). \tag{III.18}$$

Although we do not have an inductive proof valid for all orders we did the same calculation for $D^{(4)}_m(p, l_1, l_2, l_3, l_4)$, figuring in ladder-type three-loop diagrams, which indeed showed a similar behavior

$$D^{(4)}_m(p, l_1, l_2, l_3, l_4) = O\left(\frac{1}{m^6}\right) \tag{III.19}$$

in accordance with the requirement of the validity of the energy expansion. This calculation was done with a Mathematica program and attempts to calculate $D^{(6)}_m$ did not succeed because of the prohibitively large amount of CPU time required. We have unfortunately not found an inductive proof that lets us extend these results to all orders. These calculations show that in the toy model, which is free of gauge complications, the power counting theorem works correspondingly to Weinberg's power counting theorem in chiral effective QCD due to unexpected cancellations.

## IV   Lessons for general relativity

Many of the features uncovered in the last section are also applicable for general relativity. Although general relativity contains additional multigraviton vertices, the matter vertex considered in the previous section is the most dangerous. We have not been able to reformulate the general power counting formula (in harmonic gauge) such that the cancellation is explicit in all diagrams. However, we can display how the desired behavior is restored in examples which are physically relevant.

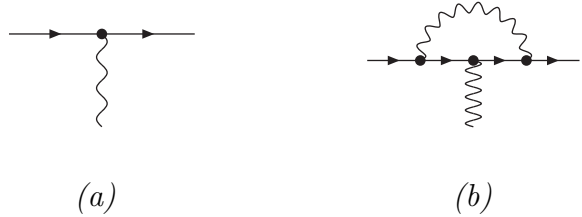

(a)　　　　　　　　(b)

Figure 10: Graviton vertex and one of the loop corrections.

Before we turn to the box diagram once again, let us consider the vertex correction, Fig. 10. Of the several contributions to the vertex at one loop only Fig. 10b and the self-energy diagrams are of the dangerous category, with two extra matter vertices and one loop.

Because the vertex coupling is the energy momentum tensor, and the energy momentum tensor is conserved, there is a nonrenormalization theorem for the matrix element at $q^2 = 0$. The general form for the vertex, consistent with the conservation $\partial^\mu T_{\mu\nu} = 0$ is

$$< p' \mid T_{\mu\nu} \mid p > = F_1(q^2) \cdot \left[ p_\mu p'_\nu + p_\nu p'_\mu - \eta_{\mu\nu} \right] + F_2(q^2) \cdot \left[ q_\mu q_\nu - \eta_{\mu\nu} q^2 \right] \quad \text{(IV.1)}$$

and at tree level $F_1 = 1$, $F_2 = 0$.

The dangerous diagrams naïvely give a correction to $F_1$ of order $\kappa^2 m^2$, and individually will do so. [Because powers of $q^2$ multiply $F_2$ it automatically is not a problem – a one loop contribution to $F_2(q^2)$ of order $\kappa^2 m^2$ is allowable in a well behaved energy expansion.] However, since $T_{\mu\nu}$ measures the physical energy and momentum we have the constraint $F_1(0) = 1$. Thus all contributions to $F_1(q^2)$ which are independent of $q^2$, in particular all corrections of order $\kappa^2 m^2$, must cancel when expressed in terms of the physical mass and momenta. This occurs by the cancellation of the vertex 10b with the renormalization due to the self energy. This is entirely analogous to the nonrenormalization of the charge form factor in QED at $q^2 = 0$.

One can also repeat the exercise to show that the sum of the two diagrams in Fig. 7a,b, but with two gravitons instead of scalars, behaves as $\kappa^2 m^2$ instead of the $\kappa^2 m^3$ behavior as given by the naïve power counting. The gravitational vertex is more complicated than the scalar one, for example involving $\kappa p_\mu (p+k)_\nu$ at a given vertex instead of $\kappa m^2$. However, in the counting of powers of mass $p_\mu k_\nu$ is already one factor fewer power of the mass than is $p_\mu p_\nu$. Therefore in showing that the $\kappa^2 m^3$ behavior is not present in the sum of diagrams, we need only consider the $\kappa p_\mu p_\nu$ portion of the vertex, which is common to all vertices and which will not upset the cancellation of the two diagrams. The proof then goes through exactly as in the last section.

Finally, we seek to demonstrate the cancellation in the box plus crossed box diagrams.

In order to show that the situation, including the necessary cancellations, is analogous to the one in the toy model, we calculate the contribution from the box and crossed box graphs on Fig. 7a,b and we will see that the cancellation of the $\geq O\left(\frac{1}{q^2}\right)$ terms *does* occur in the *real* part of the amplitude and that in the nonrelativistic limit, i. e. for small values of spatial momenta $p \ll \sqrt{s}$, the toy model gives the correct coefficients of the leading terms. This could be expected as we will explain later.

Figure 11: Propagators and the relevant vertex in harmonic gauge.

In harmonic gauge the graviton propagator is

$$\frac{i}{q^2 - i\epsilon} \cdot P_{\mu\nu,\alpha\beta} \quad \text{with} \quad P_{\mu\nu,\alpha\beta} = \eta^\alpha_{(\mu} \eta^\beta_{\nu)} - \frac{1}{2}\eta_{\mu\nu}\eta^{\alpha\beta}, \qquad \text{(IV.2)}$$

the heavy mass propagators are

$$\frac{i}{p^2 - m^2} \qquad \text{(IV.3)}$$

and the graviton-scalar-scalar vertex is

$$-i\kappa \cdot V^{(m)}_{\mu\nu}(p,p') \quad \text{with} \quad V^{(m)}_{\mu\nu}(p,p') = p_{(\mu}p'_{\nu)} - \frac{1}{2}\eta_{\mu\nu} \cdot (p \cdot p' - m^2), \tag{IV.4}$$

see also Fig. 11. Using these rules and the usual Feynman parameterization in the loop integral we arrive at

$$\mathcal{M}_{(c)box} = -\frac{\kappa^4}{64} \cdot \frac{1}{(4\pi)^{2+\frac{\epsilon}{2}}} \cdot \int_{-1}^{1} \frac{dx}{2} \int_{0}^{1} dy \cdot y(1-y) \cdot \int_{0}^{1} dz \cdot$$

$$\cdot \left\{ 4 \cdot \Gamma\left(2 - \frac{\epsilon}{2}\right) \cdot \frac{F_{(c)box}}{[M^2_{(c)box} - i\epsilon]^{2-\frac{\epsilon}{2}}} - \right. \tag{IV.5}$$

$$- 2 \cdot \Gamma\left(1 - \frac{\epsilon}{2}\right) \cdot \frac{g_{\mu\nu}F^{\mu\nu}_{(c)box}}{[M^2_{(c)box} - i\epsilon]^{1-\frac{\epsilon}{2}}} +$$

$$\left. + \Gamma\left(-\frac{\epsilon}{2}\right) \cdot \frac{(g_{\mu\nu}g_{\alpha\beta} + g_{\mu\alpha}g_{\nu\beta} + g_{\mu\beta}g_{\nu\alpha}) \cdot F^{\mu\nu\alpha\beta}_{(c)box}}{[M^2_{(c)box} - i\epsilon]^{-\frac{\epsilon}{2}}} \right\},$$

where the IR singularities (at $\epsilon \to 0$) come from the first term and the UV singularities come from the last term. The functions $F$ are determined as

$$F_{box} = \left[V^{m_1}_{\mu\nu}(p_1, p_1 + k) \cdot P^{\mu\nu,\alpha\beta} \cdot V^{m_2}_{\alpha\beta}(p_2, p_2 - k)\right] \cdot$$
$$\cdot \left[V^{m_1}_{\lambda\rho}(p_1 + k, p_1 + q) \cdot P^{\lambda\rho,\kappa\delta} \cdot V^{m_2}_{\kappa\delta}(p_2 - k, p_2 - q)\right]$$
$$F_{cbox} = \left[V^{m_1}_{\mu\nu}(p_1, p_1 + k) \cdot P^{\mu\nu,\alpha\beta} \cdot V^{m_2}_{\alpha\beta}(p_2 - q + k, p_2 - q)\right] \cdot \tag{IV.6}$$
$$\cdot \left[V^{m_1}_{\lambda\rho}(p_1 + k, p_1 + q) \cdot P^{\lambda\rho,\kappa\delta} \cdot V^{m_2}_{\kappa\delta}(p_2, p_2 - q + k)\right]$$

and

$$[M_{(c)box}]^2 = (1-x^2)(1-y)^2 \cdot \frac{q^2}{4} + y^2 \cdot [\ ]_z \tag{IV.7}$$

with

$$[\ ]_z^{box} = (1-z)^2 \cdot m_1^2 + z^2 m_2^2 - z(1-z)(s - m_1^2 - m_2^2)$$
$$[\ ]_z^{cbox} = (1-z)^2 \cdot m_1^2 + z^2 m_2^2 - z(1-z)(u - m_1^2 - m_2^2), \tag{IV.8}$$
$$u \equiv 2(m_1^2 + m_2^2) - s - q^2 \equiv u_0 - q^2.$$

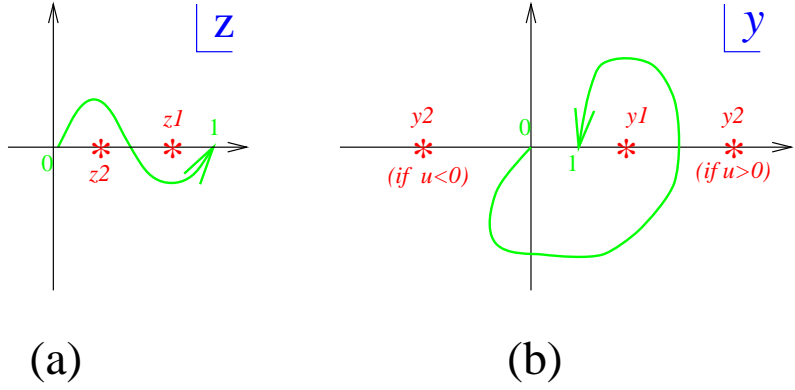

Figure 12: (a) The complex path of $z$ integration. (b) What this path becomes in the complex $y$ plane. Singularities are shown in red, the path of integration in green.

The quantity $[\ ]_z^{cbox}$ is always positive, corresponding to the fact that the crossed box diagram has no imaginary part (which would emerge from the point where $M^2_{(c)box} \to 0$). The corresponding $[\ ]_z^{box}$ does have zeroes which can be avoided by a complex continuation in $z$ as shown in Fig. 12.

After some a tedious algebra we managed to separate out the IR divergent part which only stems form the $x = \pm 1, y = 0$ corners of the integration region. The coefficients were extracted with a REDUCE program and cast in the form

$$i\mathcal{M}_{(c)box} = -\frac{\kappa^4}{8(4\pi)^2} \int_0^1 dz \int_0^{\pi/2} d\beta \cdot \cos\beta \sin\beta \cdot \left(\sum_{k=1}^4 h_k(\gamma, z)\psi_k(b)\right) \quad \text{(IV.9)}$$

with

$$\gamma = \frac{\cos\beta - \sin\beta}{\cos\beta + \sin\beta} \quad \text{(IV.10)}$$

and

$$b = \frac{q^2}{4} \cdot \cos^2\beta + [\ ]_z \cdot \sin^2\beta - i\epsilon, \quad \text{(IV.11)}$$

and

$$\psi_1(b) = \frac{1}{b}, \quad \psi_2(b) = \frac{1}{b^2}, \quad \psi_3(b) = \ln b, \quad \psi_4(b) = \frac{\ln b}{b^2}, \quad \text{(IV.12)}$$

and the functions $h_k(\gamma, z)$ turn out to be polynomials in $q^2$; they contain about 932 terms that we do not display here. Their algebraic structure can be understood though and we find that the only nonpolynomial dependences in *any* of the variables come from seven special functions of $\gamma$ and their singularities are avoided by the above prescription to use a complex path of integration for $z$.

The calculation at this point contains *no* approximation at all. However, the integrals cannot be done by quadrature. For our purposes it is only necessary to separate the terms which are divergent in the limit $q \to 0$, and due to the fact that the integration contours defined above avoid the singularities we can bring the $q \to 0$ limit into the integrands. For lack of space we do not describe all the details of the thorough analysis of the divergent terms. What we do is we first observe that cutting the $\beta$ integration in two parts, $\int_0^{\rho(s,m_1^2,m_2^2)} + \int_\rho^{\pi/2}$, the integrand in the latter is uniformly bounded and so provides no divergent terms. Using a Taylor-formula for the former part, the remainder is also seen to be uniformly bounded (this was proven for each $k$ using Weierstrass's theorem) and what remains is

$$\begin{aligned}
\mathcal{M}_{(c)box} &= -\frac{\kappa^4}{8(4\pi)^2} \cdot \Bigg\{ \frac{4}{q^2} \ln \frac{q}{2} \cdot \tilde{h}_4^{(0)} \cdot \int_0^1 \frac{dz}{[\ ]_z^{(0)}} \\
&\quad + \frac{2}{q^2} \cdot \int_0^1 \frac{dz}{[\ ]_z^{(0)}} \cdot \left( \tilde{h}_4^{(0)} + \tilde{h}_2^{(0)}(0, z) \right) \\
&\quad - \frac{i\pi}{2q} \cdot \int_0^1 \frac{dz \cdot \left( \tilde{h}_2^{(0)} \right)'(0, z)}{[\ ]_z^{(0)} \cdot \sqrt{i\epsilon - [\ ]_z^{(0)}}} \\
&\quad - \ln \frac{q}{2} \cdot \Bigg[ \int_0^1 \frac{dz}{[\ ]_z^{(0)}} \left( \tilde{h}_1^{(0)}(0, z) - 4\tilde{h}_4^{(1)} \right) + \\
&\quad + \int_0^1 \frac{dz}{\left( [\ ]_z^{(0)} \right)^2} \cdot \left( \left( \tilde{h}_2^{(0)} \right)''(0, z) - \tilde{h}_4^{(0)} \cdot \phi(z) \right) \Bigg] \Bigg\},
\end{aligned} \quad \text{(IV.13)}$$

where $[\ ]_z^{(0)} \equiv ([\ ]_z)_{q \to 0}$ and $\tilde{h}^{(0)}(x, z) \equiv h(x \to \tan \beta, z)_{q \to 0}$.

Also, $\tilde{h}_4^{(1)} \equiv \lim_{q \to 0} \frac{h_4 - h_4^{(0)}}{q^2}$ and the prime always refers to the derivative in $x$ and we used the notation

$$\phi(z) = \begin{cases} 1 + 4z(1-z) & \text{for } cbox \\ 1 & \text{for } box. \end{cases}$$

Here, $h_4^{(0)} = 8\left[s^2 - 2s(m_1^2 + m_2^2) + (m_1^4 + m_2^4)\right]^2$ is independent of $\beta, z$. This term generates the leading divergence, and as we will see this term is the only one present in the toy model of Sec. III.

The most divergent terms are multiplied by the integrals $\int_0^1 \frac{dz}{[\,]_z^{(0)}}$. A substitution of variables $z \to y$ with $\frac{1}{z} + \frac{1}{y} = 2$, has the interesting property

$$[\,]_z^{box} = \frac{[\,]_y^{(0,cbox)}}{(2y-1)^2},$$

and tracing the path of the complex $y$ integration as shown in Fig. 12 we find a residue term in

$$\int_0^1 \frac{dz}{[\,]_z^{box}} = -\int_0^1 \frac{dz}{[\,]_z^{(0,cbox)}} + \frac{i\pi}{p\sqrt{s}}, \tag{IV.14}$$

i.e. the sum of the two integrals figuring in (IV.13) is a pure residue. We find that the leading divergence is IR and UV finite:

$$\sim -\frac{i}{16\pi} \cdot \frac{\ln q}{q^2} \cdot \frac{\kappa^4}{p\sqrt{s}} \cdot \left[s^2 - 2s(m_1^2 + m_2^2) + (m_1^4 + m_2^4)\right]^2 \tag{IV.15}$$

is pure imaginary and as such can be included in an unobservable phase.

The next-to-leading, $\frac{1}{q^2}$ divergence contains the same factor and also results in a pure imaginary contribution

$$\sim -\frac{i}{16\pi} \cdot \frac{1}{q^2} \cdot \frac{\kappa^4}{p\sqrt{s}} \cdot \left[s^2 - 2s(m_1^2 + m_2^2) + (m_1^4 + m_2^4)\right]^2 \cdot \left(\frac{1}{\epsilon} + \frac{C - 1 - \ln 4\pi}{2}\right) \tag{IV.16}$$

where the $\epsilon \to 0$ singularity is of IR origin.

The ambiguity in (IV.15) in defining $\ln q \equiv \ln \frac{q}{\mu} + \ln \mu$ is a reflection of the ambiguity in the IR regulator to remove IR divergencies. Total removal is achieved only after taking into account soft graviton processes as in [4] and different choices of $\mu$ show up as different constants in (IV.16):

$$\frac{C - 1 - \ln 4\pi}{2} \longrightarrow \frac{C - 1 - \ln 4\pi}{2} + \ln \frac{\mu'}{\mu}. \tag{IV.17}$$

The $O\left(\frac{1}{q}\right)$ divergencies are much harder to calculate. We found, by a similar calculation, that the $O\left(\frac{1}{q}\right)$ part has *no* imaginary part, contains *no* IR or UV singular terms. Its coefficient is a complicated function of $s$, $m_1^2$ and $m_2^2$ which we do not display for lack of space but whose $p \to 0$ limit is

$$\sim \frac{\kappa^4 m_1^2 m_2^2}{128 q} \cdot \left[m_1 + m_2 + O\left(\frac{p}{m}\right)\right]. \tag{IV.18}$$

This term, which still contains *one* power of $\kappa m$ too much, corresponds to a contribution to the classical potential

$$V(r) \longrightarrow -G \frac{m_1 m_2}{r} \cdot \left[\ldots - \frac{G(m_1 + m_2)}{4r} + \ldots\right] \tag{IV.19}$$

and forms part of the *classical* general relativity corrections which are encoded in the one-loop diagrams of Fig. 7 when the harmonic gauge is used. We note that no similar real term arises in the toy model of Eq. (III.11).

The $p \to 0$ limit of the coefficients of $q \to 0$ divergent terms turn out to be smooth except for the imaginary part which has a $\frac{\kappa^4 m^7}{q^4 \cdot p}$ term in (IV.16). The limit we use in this calculation,

$$\frac{p}{m} \gg \frac{q}{p} \to 0 \tag{IV.20}$$

exactly corresponds to the requirement that no bound states are formed

$$\begin{aligned} G m_1 m_2 \cdot q &\sim \frac{p^2}{2m_1} + \frac{p^2}{2m_2} \\ &\text{i. e.} \\ 2 G m_1 m_2 &\ll \frac{p^2}{q \cdot m_{red}} \to \infty. \end{aligned} \tag{IV.21}$$

We finally comment on why out toy model gave the correct leading divergencies. At leading order in $\frac{q}{m}$, the heavy scalar vertex becomes

$$-\frac{i}{2}\kappa \cdot [p_\mu \cdot p'_\nu + p_\nu \cdot p'_\mu - \eta_{\mu\nu} \cdot (p \cdot p' - m^2)] \longrightarrow -i\kappa m^2 v_\mu v_\nu. \tag{IV.22}$$

This vertex is attached to a graviton propagator and is multiplied by $P^{\mu\nu}_{\alpha\beta}$ and so becomes

$$-i\kappa m^2 \cdot \left(v_\alpha v_\beta - \frac{1}{2}\eta_{\alpha\beta}\right), \tag{IV.23}$$

so when we multiply this by the heavy scalar on the other end of the graviton propagator, the full theory, as compared to the toy model, provides and extra factor

$$(v_1 \cdot v_2)^2 - \frac{1}{2} \equiv \frac{(2p_1 \cdot p_2)^2 - 2m_1^2 m_2^2}{4m_1^2 m_2^2} \equiv \frac{[s^2 - 2s(m_1^2 + m_2^2) + (m_1^4 + m_2^4)]}{4m_1^2 m_2^2}. \tag{IV.24}$$

Comparing now (IV.15) with the appropriate, $O\left(\frac{\ln q}{q^2}\right)$ part of (III.11) we see that the square of (IV.24) shows up, corresponding to the presence of *two* graviton lines. A similar comparison between (IV.18) and (III.11) shows that in the toy model the $\frac{1}{q}$ divergence is absent. It is nonzero in (IV.18) only due to subleading contributions to (IV.22) and (IV.23).

# V  Power counting in a physical gauge

In covariant gauges, both classical and quantum effects are included in the same Feynman diagram. The simplest example is the one graviton exchange diagram, which includes the classical Newton potential, but loop diagrams also exhibit this property [2]. However, in physical gauges, classical and quantum effects are separated. The physical quantum degrees of freedom are transverse and traceless [6], corresponding to massless spin-two quanta. In a multipole expansion the monopole term, which generates the Newton potential, is classical, while the spin-two degrees of freedom couple to the quadrupole term. This suggests that the dominant $\kappa m^2$ coupling that caused us do much trouble in Sec. II is to be associated with classical physics while the quantum degrees of freedom have a milder behavior. We will show that this is the case for the interaction of two heavy masses. This allows a quantum power counting which is well behaved.

Covariant gauges, in particular the harmonic gauge treated covariantly, are preferred for practical calculations [7]. When combined with the background field technique, they can explicitly retain the invariances of general relativity. In contrast the construction of the physical gauge quantum theory severely disturbs the underlying symmetry of general coordinate invariance [8]. One picks a preferred frame for the quantization, and the split of quantum and classical degrees of freedom depends on that frame. Nevertheless a physical gauge is attractive conceptually because only the physical radiation degrees of freedom are quantized. This is analogous to the Coulomb gauge quantization of QED.

Although the general metric tensor $g_{\mu\nu}$ has ten real components, the radiation field has only two independent degrees of freedom, corresponding to helicity $\pm 2$ [6]. For a wave propagating in the $z$ direction, a harmonic gauge constraint plus residual gauge freedom can be used to reduce the ten original components of a polarization tensor $\epsilon_{\mu\nu}$ such that only $\epsilon_{11} = -\epsilon_{22}$ and $\epsilon_{12} = \epsilon_{21}$ are nonvanishing (other choices are also possible.) The helicity states with $\lambda = \pm 2$ can be found from these via

$$\epsilon_\pm = \epsilon_{11} \mp i\epsilon_{12} = -\epsilon_{22} \mp i\epsilon_{12}. \tag{V.1}$$

When quantized, these become the graviton degrees of freedom.

Let us consider the interaction of two heavy masses, which have a small relative momentum. The momentum of one of the masses can be written as

$$\begin{aligned} p_\mu &= mv_\mu + \overline{p}_\mu \\ v_\mu &= (1,0,0,0)_\mu. \end{aligned} \tag{V.2}$$

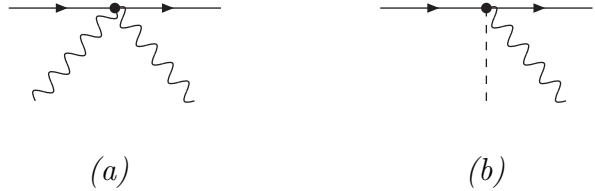

Figure 13: Suppressed matter–gravity interaction vertices.

Since

$$q_\mu = \overline{p}_\mu - \overline{p}'_\mu, \tag{V.3}$$

we will treat the residual $\overline{p}$ as of order $q$ in the energy expansion. If we quantize in the frame where $v_\mu$ defines the timelike direction, then the physical graviton degrees of freedom will only couple to the spacelike traceless components of the energy-momentum tensor $T_{ij}$. Since the large mass does not contribute to these, we know

$$T_{ij} = \overline{p}_i \overline{p}'_j + \overline{p}'_j \overline{p}_i \tag{V.4}$$

and we count $T_{ij}$ as order $q^2$, whereas $T_{00}$ is of order $m^2$. The same suppression is present for the two graviton interaction of Fig. 13a, and the coupling of one classical Newtonian field and one physical quantum field as shown in Fig. 13b. From the general vertex

$$\tau_{\lambda\eta,\rho\sigma} = i\frac{\kappa^2}{2} \cdot \left\{ I_{\eta\lambda,\alpha\delta} \cdot I^{\delta}_{\beta,\rho\sigma} \cdot \left(p^\alpha p'^\beta p'^\alpha p^\beta\right) \right.$$

$$-\frac{1}{2} \cdot (\eta_{\eta\lambda} \cdot I_{\beta\sigma,\alpha\beta} + \eta_{\rho\sigma} \cdot I_{\eta\lambda,\alpha\beta}) \cdot p'^\alpha p^\beta \quad \text{(V.5)}$$

$$\left. -\frac{1}{2}\left(I_{\eta\lambda,\rho\sigma} - \frac{1}{2}\eta_{\eta\lambda}\eta_{\rho\sigma}\right) \cdot [p\cdot' - m^2] \right\}.$$

we have

$$\begin{aligned} \tau_{00,00} &\sim \kappa^2 m^2 \\ \tau_{00,ij} &\sim \kappa^2 q^2 \\ \tau_{ij,kl} &\sim \kappa^2 q^2. \end{aligned} \quad \text{(V.6)}$$

What we see from these couplings is that the physical quantum degrees of freedom have a reduced matter coupling with two fewer powers of $m$ compared to the harmonic gauge counting rules.

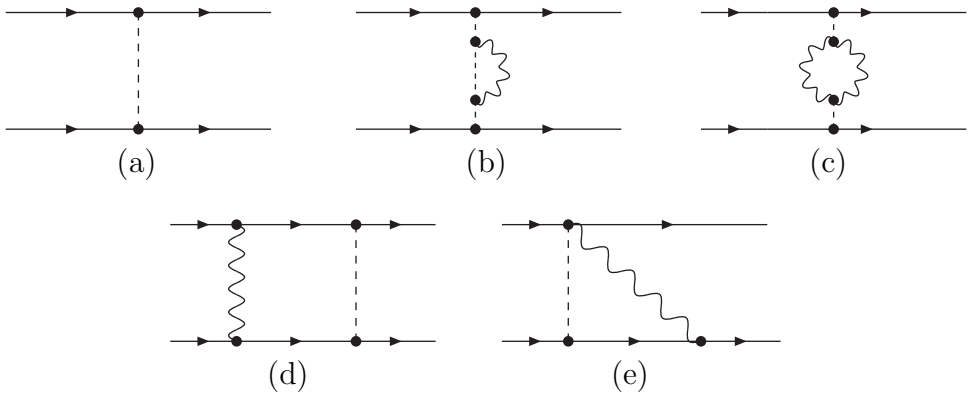

Figure 14: Interaction of two masses in a physical gauge.

Various other diagrams describing the interaction of two masses in physical gauges are shown in Fig. 14. The dashed line represents the classical Newtonian potential while the wavy line describes the quantum degrees of freedom. In the normalization used throughout this work, the classical interaction is again of order $\frac{\kappa^2 m^4}{q^2}$. Corrections due to vacuum polarization in Fig. 14b,c have the same powers of $m$ but two extra powers of $\kappa^2$, leading to a result of order

$$\frac{\kappa^2 m^4}{q^2} \cdot \kappa^2 q^2 \cdot (a + b \log q^2) \tag{V.7}$$

so that these diagrams have the expected expansion parameter $\kappa^2 q^2$ as desired. An important feature of the physical gauge is that diagrams with extra matter couplings are suppressed. For example, the mixed box diagram of Fig. 14d has two factors of $\frac{1}{m}$ from the propagators, but no compensating factors of the mass in the vertex coupling or the physical gravitons. This means that this diagram is suppressed by $\frac{1}{m^2}$ compared to the vacuum polarization corrections. All of the diagrams with graviton–matter vertices are suppressed by powers of $\frac{1}{m}$. This leaves the vacuum polarization diagram as the leading quantum correction is this gauge, with a well behaved expansion parameter.

The above results may be seen more easily in different normalization for the fields, typical for the nonrelativistic limit. Here, we divide all matter vertices by a factor of $2m$, so that $T_{00} = m$ for a particle at rest. In this normalization (used also in Heavy Quark Effective Theory [9]) propagators have no factor of $\frac{1}{m}$. However, the coupling of the transverse traceless degrees of freedom to matter fields are proportional to

$$T_{ij} = \frac{\overline{p}_i \overline{p}'_j + \overline{p}'_i \overline{p}_j}{2m} \tag{V.8}$$

and vanish as $m \to 0$. Then all diagrams with matter coupling of graviton simply drop out, leaving only the vacuum polarization diagram in this gauge.

To close this section, we note that the use of a physical gauge seems to be required if we are to be able to introduce the idea of a purely classical source. In the harmonic gauge, the inclusion of the matter propagator was required to properly identify the classical and quantum corrections. Both vertex and vacuum polarization diagrams are important. One could not at the start of the calculation take the mass to infinity and treat this resulting field in the classical limit. The reason is that the vertex coupling strength also goes to infinity in this limit. However, in the physical gauge, the diagrams with matter propagators and couplings to transverse quantum fields are unimportant in the limit $m \to \infty$. By taking this limit in this gauge, one obtains a classical source, with an interaction which receives quantum corrections.

# VI    Summary

We were motivated for this study by the observation of a class of diagrams which in harmonic gauge would apparently upset the utility of the energy expansion, and indeed also spoil the classical limit of the theory. Part of the problem is due to the fact that the graviton propagator in harmonic gauge includes both classical and quantum effects. By consideration of several diagrams we were able to demonstrate the nature of the cancellations in harmonic gauge which removed the undesirable expansion parameter and led to a well defined energy expansion. The logic for this behavior is clearer in a physical gauge, analogous to Coulomb gauge in QED, even if explicit calculations are much more painful in such a gauge. The physical transverse traceless quantum degrees of freedom only couple with a reduced strength in problems with nearly static matter fields. In the limit that $m$ becomes very large, the effect of matter couplings become negligible, and the modification to the gravitational self-interactions (i. e. vacuum polarization) become the most important quantum corrections. These self-interactions of the sector satisfy the Weinberg power counting theorem without any problems. This indicates that the quantum energy expansion is well behaved in physical gauges, and hence by extension in all gauges as long as one is calculating gauge invariant quantities.